# From Dots to Stripes to Sheets – Shape Control of Lead Sulfide Nanostructures

Thomas Bielewicz, Mohammad Mehdi Ramin Moayed, Vera Lebedeva,

Christian Strelow, Angelique Rieckmann, Christian Klinke*

*Institute of Physical Chemistry, University of Hamburg,*

*Grindelallee 117, 20146 Hamburg, Germany*

**Abstract**

*Controlling anisotropy in nanostructures is a challenging but rewarding task since confinement in one or more dimensions influences the physical and chemical properties of the items decisively. In particular, semiconducting nanostructures can be tailored to gain optimized properties to work as transistors or absorber material in solar cells. We demonstrate that the shape of colloidal lead sulfide nanostructures can be tuned from spheres to stripes to sheets by means of the precursor concentrations, the concentration of a chloroalkane co-ligand and the synthesis temperature. All final structures still possess at least one dimension in confinement. Electrical transport measurements complement the findings.*

* E-mail: klinke@chemie.uni-hamburg.de

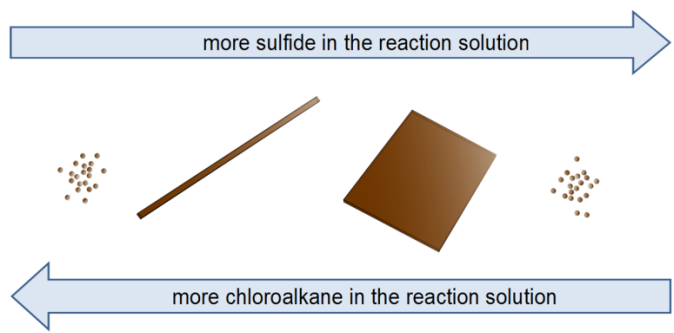

A new route to shape colloidal PbS nanomaterials is demonstrated. The structures cover all dimensionalities from 0D to 3D. Additionally, the effect of temperature on the shape and thickness of PbS nanosheets is shown and electrical transport measurements complement the findings.

## Introduction

By influencing the shape and size of nanocrystals it is possible to also influence their chemical and physical properties.[1,2,3] These depend strongly on the structures' dimensionality. The optical and electrical properties of a 0D material (e.g. quantum dots) differ strongly from the ones of a 1D nanomaterial (e.g. nanowires). While first possess discrete energy levels,[4] latter possess so-called van-Hove singularities in the density of states.[5] Nowadays, also two-dimensional materials (2D) gain attention due to good conductivity in the plane while still being tunable in their electronic properties due to confinement in height.[6,7,8,9,10,11,12,13] Colloidal chemistry represents a promising route to synthesize semiconductor nanomaterials which are inexpensive and easy processable as thin-films for electronic devices.[14,15,16,17,18,19,20,21] This method allows for shape control by changing e.g. the synthesis temperature or ligand concentration.[22,23,24,25]

To obtain anisotropic nanostructures crystal growth must be prevented in certain directions which is well achievable in PbS,[26,27] despite PbS having a cubic symmetry. To alter the growth of PbS to obtain different kinds of anisotropic structures pressure and temperature can be enough to transform already grown PbS nanowires to two-dimensional structures.[28] Another convenient way is to introduce a chloroalkane to a colloidal nanoparticle synthesis which can yield a lot of different structures depending on other parameters in the synthesis.[29,30,31] By confinement in height, producing two-dimensional sheet-like structures, the effective band gap can be tuned from the bulk value of 0.41 eV[32] up to values of over 1 eV while charge carriers can still move freely in the sheets' plane. The rather large Bohr exciton radius of PbS of about 18 nm[33] also helps to shift the absorption wavelength from the bulk value of ~3000 nm to the visible range.[34] This makes PbS an interesting semiconductor material for applications in infra-red photodetectors[35] and solar cells.[36,37] Moreover, its capability for carrier multiplication attracted considerable attention.[38,39]

In our first findings on two-dimensional PbS *via* colloidal synthesis,[40] the addition of a chloride compound was crucial because it influenced the kinetics of nucleation and growth in such a way that ultra-small PbS particles (< 3 nm) merged into larger sheets by oriented attachment. The driving force of this attachment is the reduction of the area of highly reactive {110} facets of the ultra-small particles. At the same time a vertical growth is prohibited by a self-assembly of oleic acid on the {100} facets of the crystals. Later, we showed that through a variation of the oleic acid concentration the sheets could be controlled in their thickness and sheets with heights ranging from 4 up to 20 nm could be synthesized.[18] At the same time, the group of Sun *et al.* showed that using different chloroalkanes at different temperatures the thickness could also be controlled from 2 up to 4.6 nm.[41] Here, we show that the thickness of the sheets is only dependent on the reaction temperature regardless which chloroalkane is used. We also show how the shape of PbS nanostructures can be controlled by changing the molar ratio between the lead precursor and the sulfur source, but also by changing the concentration of the essential chloroalkane co-ligand. Latter allows tuning the shape from spherical nanoparticles to stripes and to sheets. It is astonishing that small changes can lead to a variety of new PbS structures, in particular elongated PbS stripes with a length of several tens of microns. Nanostripes can be used in further studies because of their ease of processing. We show that they possess even better electronic properties compared to what we found previously for PbS sheets.[42]

In syntheses of colloidal chalcogenide nanomaterials the molar ratio between the precursors usually varies between 1:5 to 5:1.[43] In our first reported synthesis of ultra-thin lead sulfide nanosheets the ratio between the lead and sulfur source was already at an unusual high molar ratio of 14:1.[40] Basis for the synthesis is a standard procedure for spherical nanoparticles, which was then complemented with 1,1,2-trichloroalkane (TCE) to produce nanosheets. Both approaches contain tri-n-octyl phosphine (TOP) as a co-ligand. Recently, we found that increasing the molar ratio between lead and sulfur to 120:1 makes it possible to obtain nanosheets without any TOP in the reaction solution (with spherical nanoparticles as by-product).[18] In studies presented here, we demonstrate that under appropriate conditions TOP is no longer necessary to produce large, thin and smooth nanosheets or nanostripes of lead sulfide. By increasing the molar ratio even beyond 120:1 (Pb:S) less ligands are needed to control the shape which could be also interesting for other colloidal syntheses where higher ratios could also lead to new shapes.

**Results and Discussion**

*Sulfide concentration*

Chloroalkanes function as weak ligands and support the formation of two-dimensional nanosheets by oriented attachment. In a previous publication, we could show that chloroalkanes possess the highest binding energy on {110} facets of the PbS crystal where lead and sulfur ions are aligned in rows.[18] These facets are also the ones on which oriented attachment takes place. Due to the adsorption of the chloroalkanes and displacement of oleic acid (OA) on these facets, they become richer in energy. Since the binding of the chloroalkanes is weaker to those facets than that of oleic acid or oleate, oriented attachment takes place to minimize the surface energy of these crystal facets.

In a first series of experiments the sulfide concentration was varied while fixing all other parameters. Using three different chloroalkanes as co-ligands, namely 1,1,2-trichloroethane (1,1,2-TCE), 2,3-dichlorobutane (2,3-DCB), and 1,3-dichlorobutane (1,3-DCB), shows that with decreasing sulfide concentration the structures' shape changes. Despite their structural differences (but similar boiling points), all three chloroalkanes yield comparable shapes at the same Pb:S molar ratios: from two-dimensional nanosheets (Pb:S ratio starting from 35:1) to long stripe-like structures and finally to monodisperse spherical nanoparticles (Pb:S ratio of 550:1) as shown in Figure 1. Not shown are high sulfide concentrations (Pb:S ratio > 35:1) where spherical nanoparticles are formed, too. It seems that only the functional group plays a major role. With a very low sulfide concentration (molar Pb:S ratio more than 500:1) only few nuclei are formed at the start of the reaction. These nuclei cannot find other nuclei easily to merge to nanosheets *via* oriented attachment but eventually stripes can still form through oriented attachment. By increasing the concentration (at molar Pb:S ratios between 300:1 and 500:1) enough nuclei are formed and oriented attachment takes place to yield sheet-like two-dimensional structures. When the sulfide concentration is increased, these stripes can grow together to form square-like sheets (molar ratios below 300:1) because there is still enough sulfide monomer present in the reaction solution. Otherwise, the stripes are stabilized too strongly by the ligands. Kiran *et al.* identified a $(PbS)_{32}$ cluster as the first stable crystal and could show by atomic force microscopy (AFM) measurements that the first larger and detectable structures are built from these clusters.[44] It is interesting, that four of those clusters do not always merge to square-like structures but can also form stripe-like structures where all four clusters merge *via* the same facet type. With a smaller sulfide concentration and when ligands are present in the reaction solution, the stripe-like structures seem to be the preferred shape of lead sulfide nanocrystals. This holds true for all considered chloroalkanes co-ligands, with just a slight variation of when which shape is obtained. In Figure 2, HRTEM and SAED of a stripe show that they are single-crystals and the borders of the crystal consists of {110} planes which confirms that stripes also grow in the same manner as PbS nanosheets.[40]

*Chloroalkane concentration*

The chloroalkanes function as co-ligands on the various crystal facets. It seems that they are mandatory for the PbS nanosheet synthesis. The sulfide concentration study showed that all three chloroalkanes yield very similar products. Thus, we take one of those chloroalkanes (2,3-DCB) and a structurally very different one, 1-chlorotetradecane (1-CTD), to perform experiments varying the concentration of the chloroalkanes (Figure 3). While with smaller amounts of chloroalkane sheet-like structures are formed, larger amounts lead to stripe-like structures and eventually to spherical nanoparticles. Omitting the chloroalkanes leads to mostly spherical nanoparticles. Thus, the chloroalkanes play a decisive role in the shape formation of the product. It is remarkable that even long-chained chloroalkanes such as 1-CTD behave the same way as the other short ones at the same reaction temperature of 135 °C.

With 1-CTD it is also possible to conduct a reaction with 1-CTD as sole solvent, omitting the original solvent diphenyl ether (DPE) completely. This reaction yielded spherical nanoparticles only. This supports our understanding that chloroalkanes function as a stabilizing agent for nanocrystals and such very high chloroalkane concentrations (in this case nearly four times the oleic acid molar amount) coordinate the crystals efficiently that no oriented attachment can take place.

One could argue that the shape evolution is just a matter of reaction volume and not the chloroalkane concentration. To make sure the chloroalkanes play a role in the lead sulfide nanoparticle formation two reactions have been performed without the chloroalkanes but with 0.2 mL and 2 mL of diphenyl ether instead, which has been injected at the same time as the chloroalkanes would have been. The outcome of these experiments shows that no two-dimensional nanosheets are formed when the chlorine source is not present (Figure S1).

It is worth to point out that PbS sheets and stripes, obtained by the aforementioned methods, manifest comparable semiconducting behaviors. We fabricated field-effect transistors (FETs) using individual PbS stripes, in order to compare their electrical properties with the formerly investigated sheets.[42] As depicted in Figure 4, application of a back-gate voltage can modulate the electrical conductivity of the channel significantly. The measurements reveal that the stripes in contact with Au electrodes show p-type behavior, indicated by a decrease of the current with increasing gate voltage. This means that the holes are the majority charge carriers in the device. A similar behavior was also observed in quantum dot FETs.[43] This similarity in type of the majority carriers is originated from the analogy in their structural properties (crystal structure, thickness, *etc*.). Therefore, the bandgaps of both PbS shapes, sheets and stripes, are equally affected by quantum confinement.[42,18,19] Moreover, using stripes with the same thickness as squared nanosheets as the active material improves the switching behavior. The on/off ratio for such devices can reach to over 3400 at room temperature, while the maximum field-effect mobility of the carriers is calculated to be up to 5.59 $cm^2V^{-1}s^{-1}$. The lateral confinement of the active channel may affect the performance of the FETs. In particular it reduces the current flowing in the "off" state, leading to higher on/off ratios and field-effect mobility.

*Effect of temperature*

Bhandari *et al.* recently reported on the synthesis of PbS nanosheets. They were able to tune the thickness of the nanosheets by using different chloroalkanes at different temperatures.[41] We believe that the thickness is a matter of the temperature only and is largely independent of the nature of the chloroalkanes. To further study this question a Pb:S ratio of 120:1 was chosen since this yielded the smoothest and largest nanosheets up to now. The syntheses were performed at 100, 120, 160, and 200 °C, respectively. The chloroalkane used for all the different temperatures has to be the same to show that only the temperature has the major effect on the thickness of the sheets. We used TCE and 1-CTD in the temperature study because their structure and thus, their chemical and physical properties are most different but the results from both chloroalkanes are basically the same, as we will show.

The first apparent difference in the reaction is the time the reaction solution took to turn black after the injection of the sulfur source, as a sign for PbS formation. While at 100 °C it takes about four minutes for the reaction solution to turn completely black, the time shortens to around 3 minutes at 120 °C and only 20 seconds at 160 °C. At 200 °C the solution turns black instantly after injection of the sulfur source.

The received products at the different temperatures vary significantly in shape. At 100 °C the sheets are very thin while being close to 1 µm in lateral dimensions (Figure 5). AFM measurements show the thickness to be around 5 nm including the ligand sphere. With only 20 °C more the nanosheets are larger in their lateral dimensions, but also in height. This is observable by the more pronounced vertical growth over a larger area of the sheets compared to a reaction temperature of 100 °C. In our previous findings where we varied the height of the sheets by changing the oleic acid concentration, the "second layer" did not grow to the borders of the nanosheets, even with the thickest sheets.[18] There, all reactions were performed at the same reaction temperature of 130 °C.

AFM shows that the thickness of the nanosheets synthesized at 160 °C is higher than the ones obtained at lower reaction temperatures. The surface of these nanosheets is also smoother as the vertical growth happened uniformly on the whole sheet (a calculated *root-mean-square* value over 3500 thickness data points on the nanosheet results in 0.1 nm). The lateral dimensions increased again while at the same time also more spherical nanoparticles were produced in parallel. At elevated temperatures the reactivity and decomposition of TAA was increased in such a way, that there is a quite high sulfide monomer concentration. Spherical particle growth also takes place from the beginning on while at the same time the lead oleate complex bonds are weakened which should also favor a spherical particle growth as the lead complex is more reactive. The amount of nanoparticles can be reduced post-synthetically by washing the product more often with toluene until the supernatant becomes clear in color. This procedure has no visible negative influence on the nanosheets, as there could be increased stacking or breaking apart.

At 200 °C the sheets become more square-like and there are less spherical nanoparticles in the product than at 160 °C. TAA decomposes faster at this temperature so there are more nuclei formed and less sulfide monomer is available in the growth phase – both favoring the kinetic product only. The product exhibits holes in the structures. Some structures are even more degenerated and do not resemble a sheet anymore. Besides the large almost quadratic holes, the borders are much rounder than the ones of the nanosheets received at 160 °C. The borders also show stronger contrast in the TEM images compared to the center of the nanosheets. AFM measurements prove that the border is the thickest part of the structures while there is not much change in thickness over the whole cross-section of nanosheets synthesized at 160 °C in comparison.

The holes and rounded borders could be due to a decomposition process of the nanosheets. The decomposition could be induced by acetate/oleic acid, as acids can etch lead sulfide nanoparticles.[46] We remove the acetate as acetic acid during the degassing step of the reaction. Anyhow, when the sulfide source TAA decomposes it forms acetate which could initiate the etching. The two-dimensional structures would probably start to decompose especially at their thinnest spots and already existing small holes. We observed such holes in sheets only at temperatures higher than 160 °C. By decreasing the TAA amount the decomposition at higher temperatures can be prevented which is shown in Figure S2 where a 550:1 Pb:S ratio was chosen and the sheets show no holes at a reaction temperature of 200 °C anymore.

In order to track possible loss or decomposition of TCE, NMR was performed (Figure S3). Despite TCE's boiling point of 113 °C all proton signals from TCE could be detected and their integrals are in good agreement with the molecular structure of TCE. At the same time no new signals were observed which implies that TCE is still present in the reaction solution well above its boiling point.

The PbS nanostructures are stabilized by a self-assembled monolayer of oleate with a height of 1.8 nm.[47] Considering the thinnest nanosheets synthesized at 100 °C, subtracting twice the height of the ligands (for top and bottom) leads to a height of the inorganic core part of about 2.4 nm which corresponds to only 4 times the lattice constant. We calculated the mean sheet thicknesses obtained at the different reaction temperatures (Figure S5) by fitting the (200) XRD peak and using Scherrer's equation with a form factor of 1.[49] The results for both, TCE and 1-CTD, are shown in Figure 6. A dependence of the sheet thickness with the reaction temperature is apparent, whereas the type of chloroalkane plays a minor role.

## Conclusion

By varying the concentration of the sulfur source in the PbS synthesis it was possible to change the shape of the product. Depending on the chosen concentrations formed nuclei at the beginning of the reaction can change to spherical nanoparticles, elongated stripes or large sheets. Starting from high molar ratios between lead and sulfur (550:1) a reaction at 135 °C takes place and yields elongated stripe like structures. Through HRTEM, we could show that the borders of the stripes are {110} planes and the growth occurs *via* oriented attachment. By reducing the molar ratio more square-like sheets are formed with all chloroalkanes used in this study.

By changing the chloroalkane amount the same transition between spherical nanoparticles to stripes and sheets takes place. Changing the chloroalkane amount has the same effect on the shape of the PbS sheets as changing the sulfur amount. This behavior is the same for all chloroalkanes used in this study independent of the molecules' structure, making reaction temperature and the concentrations of the chloroalkane the crucial parameters for shape change of the PbS structures. Thus, less toxic and cheaper chlorine compounds can be used to fabricate 2D PbS nanosheets.

By changing the reaction temperature we could drive the synthesis to obtain larger nanosheets with a flat and smooth surface over several microns while at the same time the thickness also increases with temperature from around 2 nm at 100 °C to 10 nm at 200 °C. The thinnest sheets could be good candidates for PbS solar cells.[14] In principle, the synthesis is scalable, but a protocol for the recycling of lead oleate should be developed due to the high precursor concentration.

## Methods

*Synthesis*

Lead(II) acetate tri-hydrate (Aldrich, 99.999 %), thioacetamide (Sigma-Aldrich, >= 99.0 %), diphenyl ether (Aldrich, 99 %+), dimethyl formamide (Sigma-Aldrich, 99.8 % anhydrous), oleic acid (Aldrich, 90 %), 1,1,2-trichloroethane (Aldrich, 96 %), 1,3-dichlorobutane (Aldrich, 99 %), 1-chlorotetradecane (Aldrich, 98 %) and 2,3-dichlorobutane (Acros, 98 %) were all used as received.

In a typical synthesis a three neck 50 mL flask was used with a condenser, septum and thermocouple. 860 mg of lead acetate tri-hydrate (2.3 mmol) were dissolved in 10 mL of diphenyl ether and 3.5 mL of oleic acid (OA, 10 mmol) and heated to 75 °C until the solution turned clear. Then vacuum was applied to transform the lead acetate into lead oleate and to remove the acetic acid in the same step. The solution was heated under nitrogen flow to the desired reaction temperature between 100°C and 200 °C) while at 100 °C the chloroalkane (volume depending on the desired shape) was added under reflux to the solution. After 12 minutes 0.05 – 2 mL of a 0.04 g thioacetamide (TAA, 0.5 mmol) in 6.5 mL dimethyl formamide (DMF) solution was added to the reaction solution. After 5 minutes the heat source was removed and the solution was left to cool down below 60 °C which took approximately 15 - 30 minutes depending on the reaction temperature. Afterwards, it was centrifuged at 4000 rpm for 3 minutes. The precipitant was washed two times in toluene before the product was finally suspended in toluene again for storage.

*TEM*

The TEM samples were prepared by diluting the nanosheet suspension with toluene and then drop casting 10 µL of the suspension on a TEM copper grid coated with a carbon film. Standard images were done on a JEOL-1011 with a thermal emitter operated at an acceleration voltage of 100 kV. HRTEM images were done on a JEOL JEM 2200FS (UHR) equipped with a field emitter, CESCOR and CETCOR correctors at an acceleration voltage of 200 kV.

*XRD*

X-ray diffraction measurements were performed on a Philips X'Pert System with a Bragg-Brentano geometry and a copper anode with a X-ray wavelength of 0.154 nm. The samples were measured by drop-casting the suspended nanosheets on a <911> or <711> grown silicon substrate.

*AFM*

Atomic force microscopy measurements were performed in tapping mode on a Veeco MultiMode NanoScope 3A and a JPK Nano Wizard 3 AFM in contact mode. The samples were prepared by spin-coating the nanosheet suspension on a silicon wafer.

*Device preparations and characterization*

PbS nanostripes with a length of serveral micrometers, a width of about 100 nm, and a height of about 15 nm suspended in toluene were spin-coated on silicon wafers with 300 nm thermal silicon oxide as gate dielectric. The highly doped silicon was used as backgate. The individual nanosheets were contacted by e-beam lithography followed by thermal evaporation of gold and lift-off. Immediately after device fabrication the devices were transferred to a probe station (Lakeshore-Desert) connected to a semiconductor parameter analyzer (Agilent B1500a). The transfer and output characteristics have been performed in vacuum at room temperature.

**Acknowledgments**

The authors thank the German Research Foundation DFG for financial support in the frame of the Cluster of Excellence "Center of ultrafast imaging CUI" and for granting the project KL 1453/9-1. The European Research Council is acknowledged for funding an ERC Starting Grant (Project: 2D-SYNETRA (304980), Seventh Framework Program FP7).

**Supporting Information**

The Supporting Information is available free of charge on the ACS Publications website at DOI: 10.1021/acs.chemmater.xxx.

Additional TEM images showing the products of the syntheses, NMR spectra of the ligands, and XRD patterns of the nanosheets at different synthesis temperatures are (PDF).

**Figures**

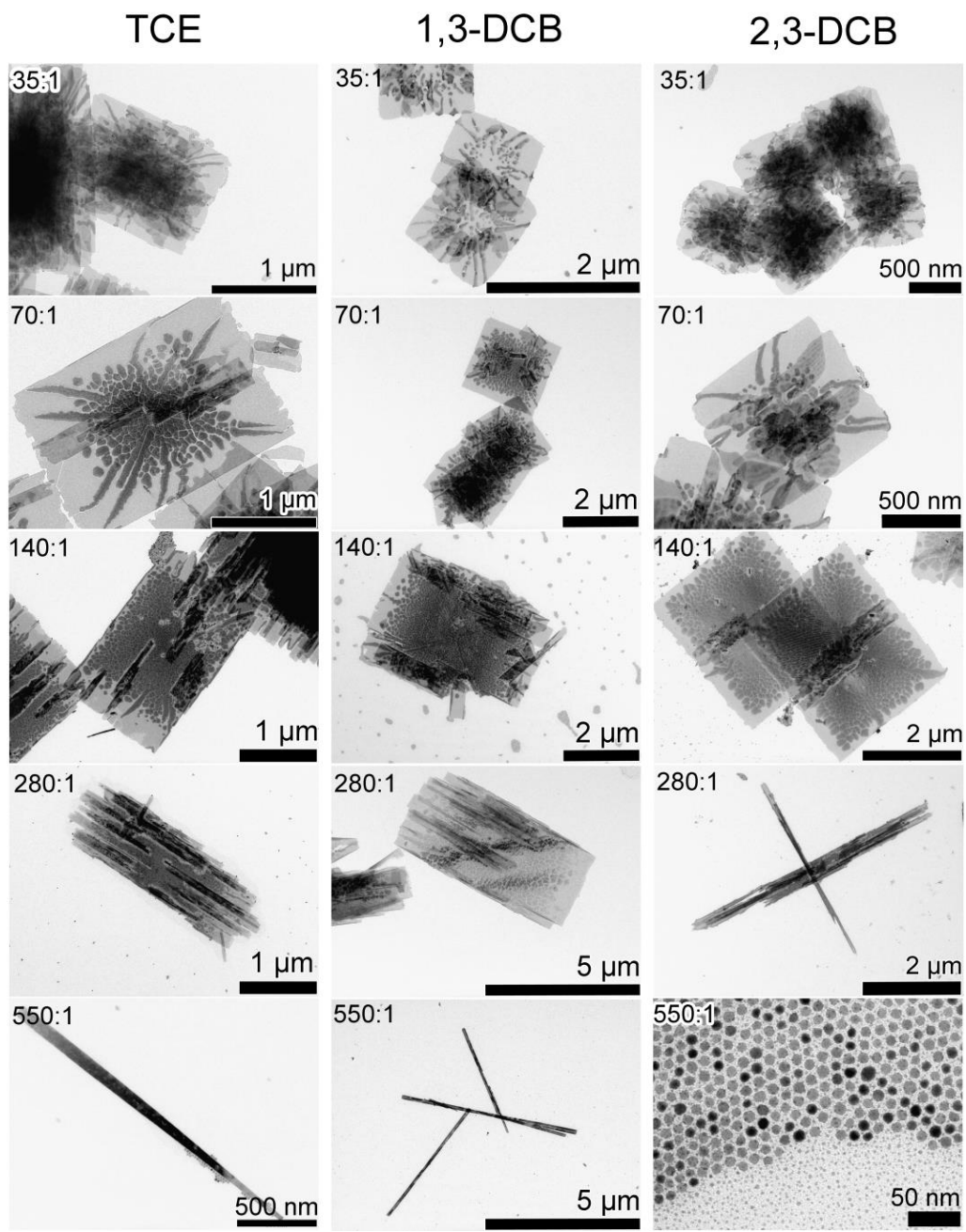

**Figure 1: PbS nanoparticle shape evolution as a function of sulfur source concentration.** *Three different chloroalkanes were used at the same reaction temperature of 135 °C. The numbers in the images correspond to the molar ratio of Pb:S. A shape evolution can be seen with all chloroalkanes. The results are basically the same for the different chloroalkanes. Only 2,3-DCB leads earlier to stripes than the other two but the shape is comparable with the others. The amount of each chloroalkane used was 0.3 mL.*

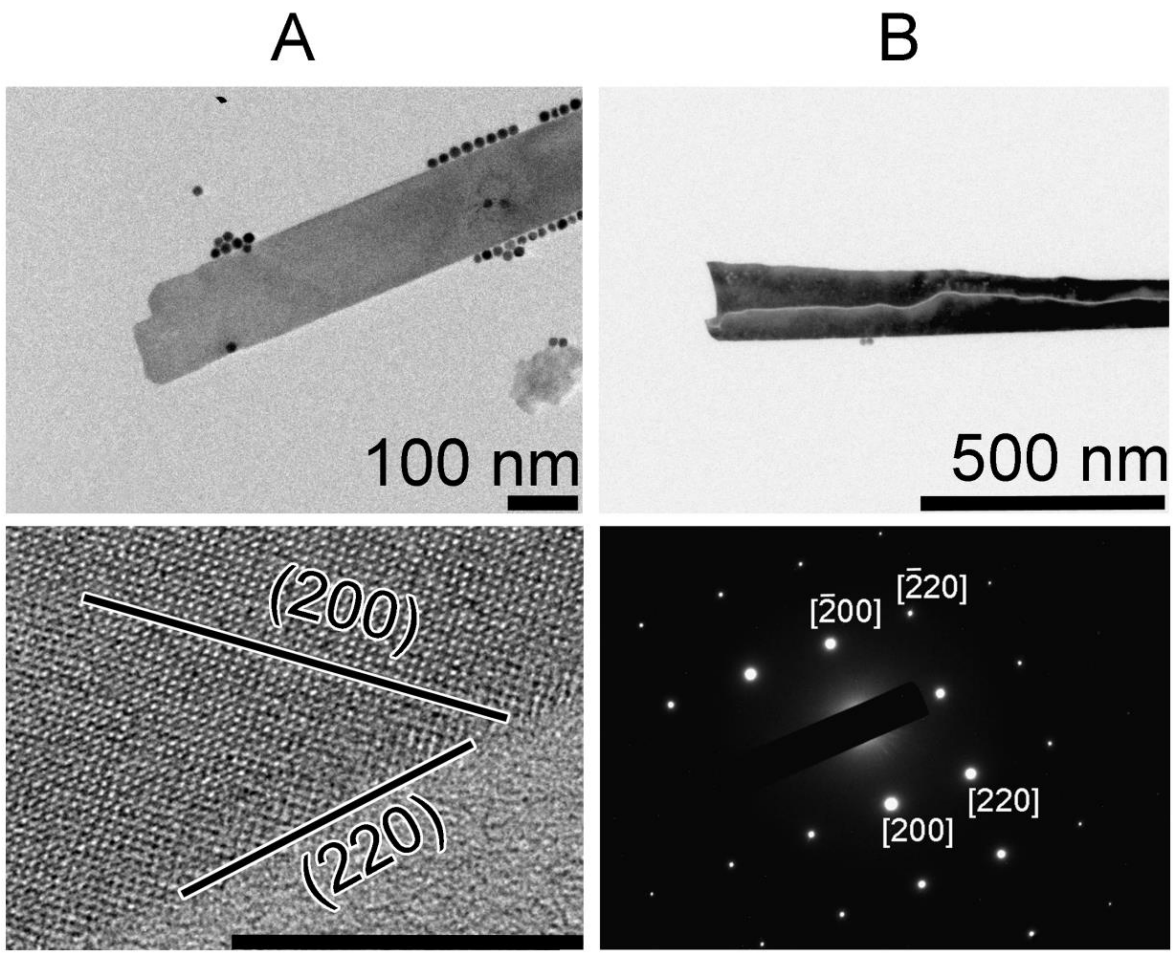

**Figure 2: HRTEM and SAED of PbS stripes.** *(A) HRTEM images show a PbS stripe synthesized at 135 °C with a molar ratio of 550:1 and 0.3 mL TCE as co-ligand shown on top. The bottom image of **A** shows the crystal structure of the same stripe (the scale bar corresponds to a length of 10 nm.) The calculated d-spacing measured in the image (d = 0.3 nm) is in good agreement with the (200) spacing from literature which is 0.2969 nm (JCPDS 5-592: galena PbS). To confirm that the planar surface of the structure is the (001) plane a SAED pattern of a stripe was performed which can be seen in **B**. By dividing the distances from the transmitting beam in the middle in the [220] direction by the distance from the middle in the [200] direction a value of 1.406 was calculated which is in very good agreement with the theoretical value of a face centered cubic (fcc) crystal of 1.414.[45]*

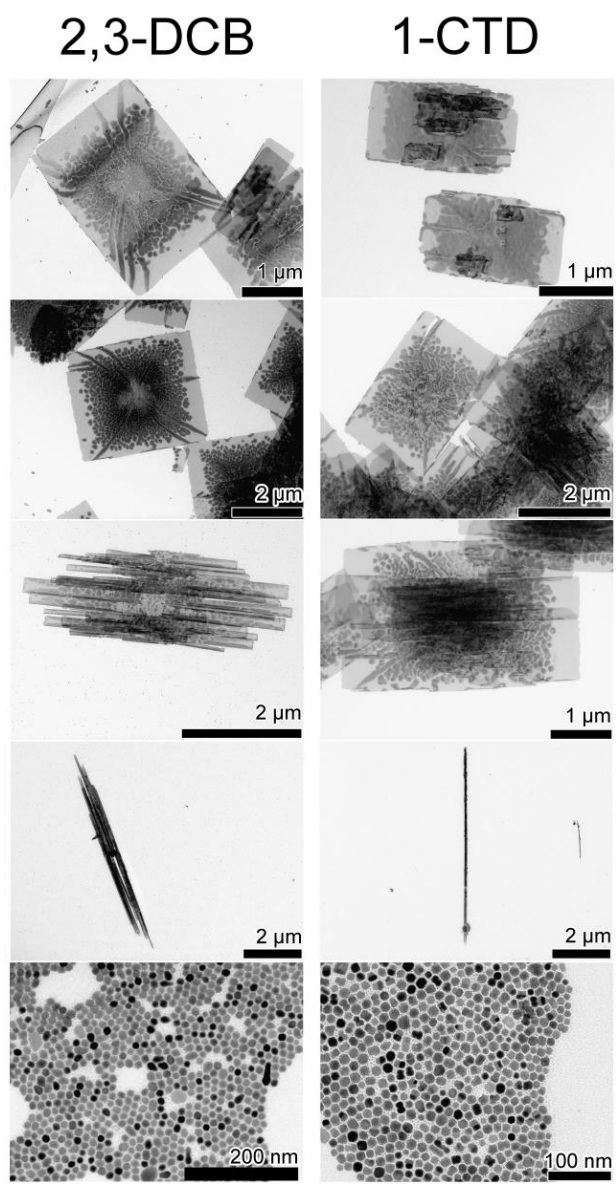

**Figure 3: PbS nanoparticle shape evolution as function of chloroalkane concentration.** *On the left the chloroalkane 2,3-DCB was used to change the shape. The volume from top to bottom is 0.15 mL, 0.3 mL, 0.6 mL, 0.9 mL and 1.25 mL, respectively. The molar ratio between Pb:S was fixed to 120:1 and the reaction temperature was 135 °C. Here, the shape of the sheets evolves from sheets to stripes to spherical nanoparticles. On the right PbS nanosheet reaction with changing 1-CTD volume at a synthesis temperature of 135 °C and a Pb:S ratio of 120:1. From top to bottom the volume is 0.4 mL, 0.8 mL, 1.6 mL, 4 mL and 10 ml. The shape of the PbS nanoparticles changes from spherical nanoparticles to stripes and finally to sheets.*

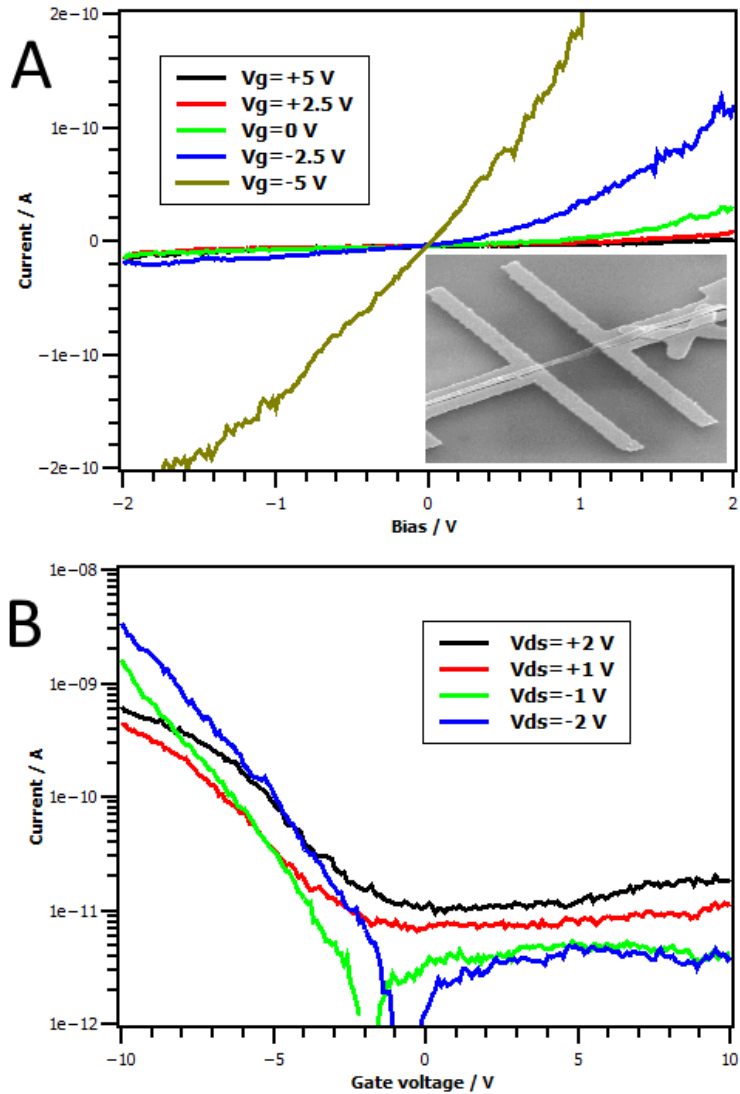

**Figure 4: Electrical transport measurements on individual PbS stripes.** *A Room temperature output characteristics of the field effect transistors using a PbS stripe as channel. The inset is the SEM image of the FET with an individual PbS stripe contacted with Au electrodes. B Transfer characteristics of the same PbS nanostripe FET. A switching effect and a p-type behavior are observed. This implies that holes are the majority carriers in these devices. The strong current changes at negative gate voltages can be understood as high field effect mobility of the carriers (holes).*

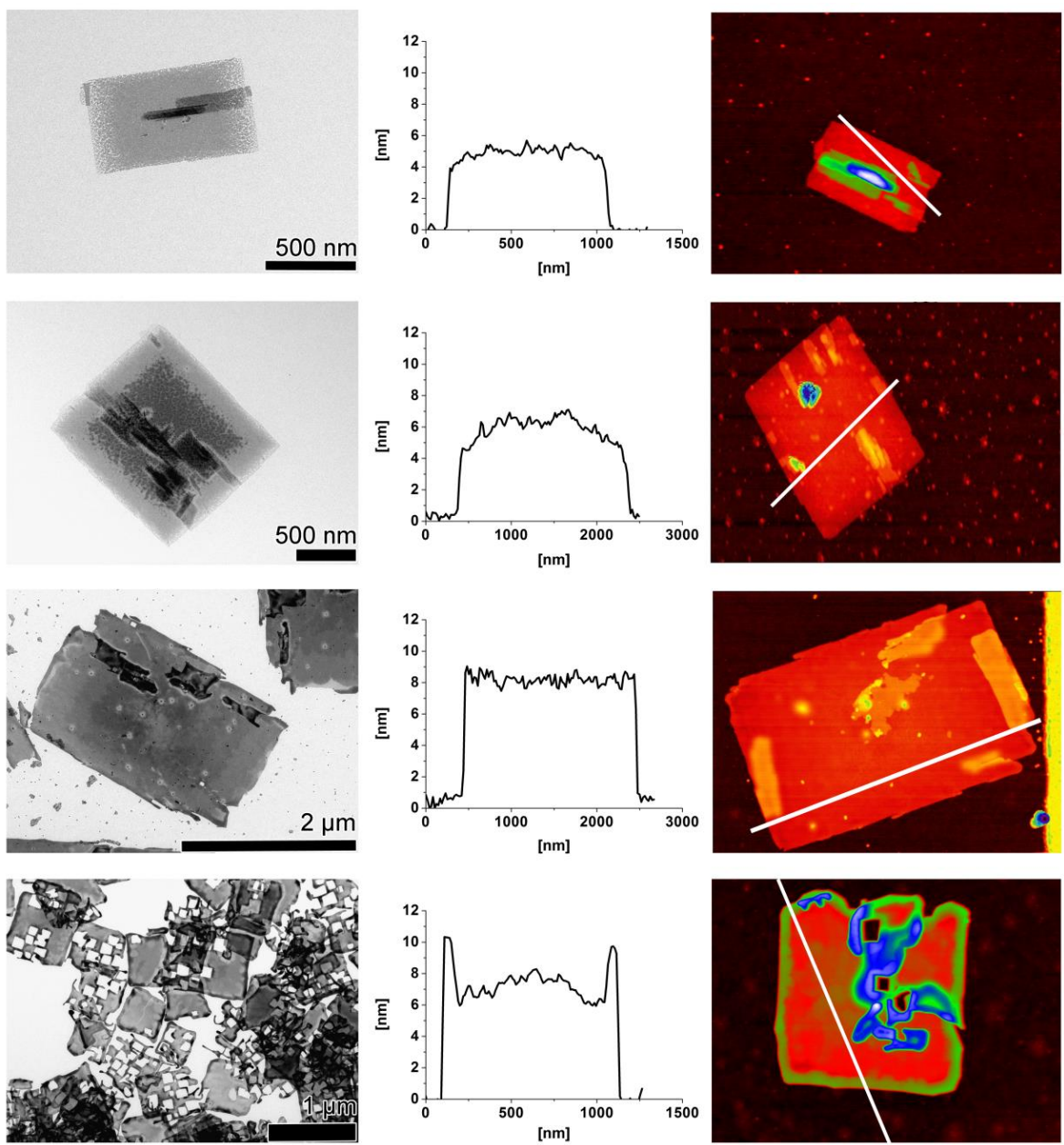

**Figure 5: Shape and thickness evolution with temperature.** *All parameters are constant and only the temperature was varied. Shown are the TEM pictures of the reaction product with TCE (0.7 mL) as co-ligand. The temperature increases from top to bottom from 100 °C, to 120 °C, to 160 °C, and finally to 200 °C. Sheets at 200 °C show a holey structure and become smaller in lateral dimensions. AFM clearly shows an increase in thickness with higher temperature while at 200 °C the borders become the thickest part of the sheet. For 1-CTD TEM images see Fig. S4 (100 °C, 200 °C) and Fig. 3(135 °C).*

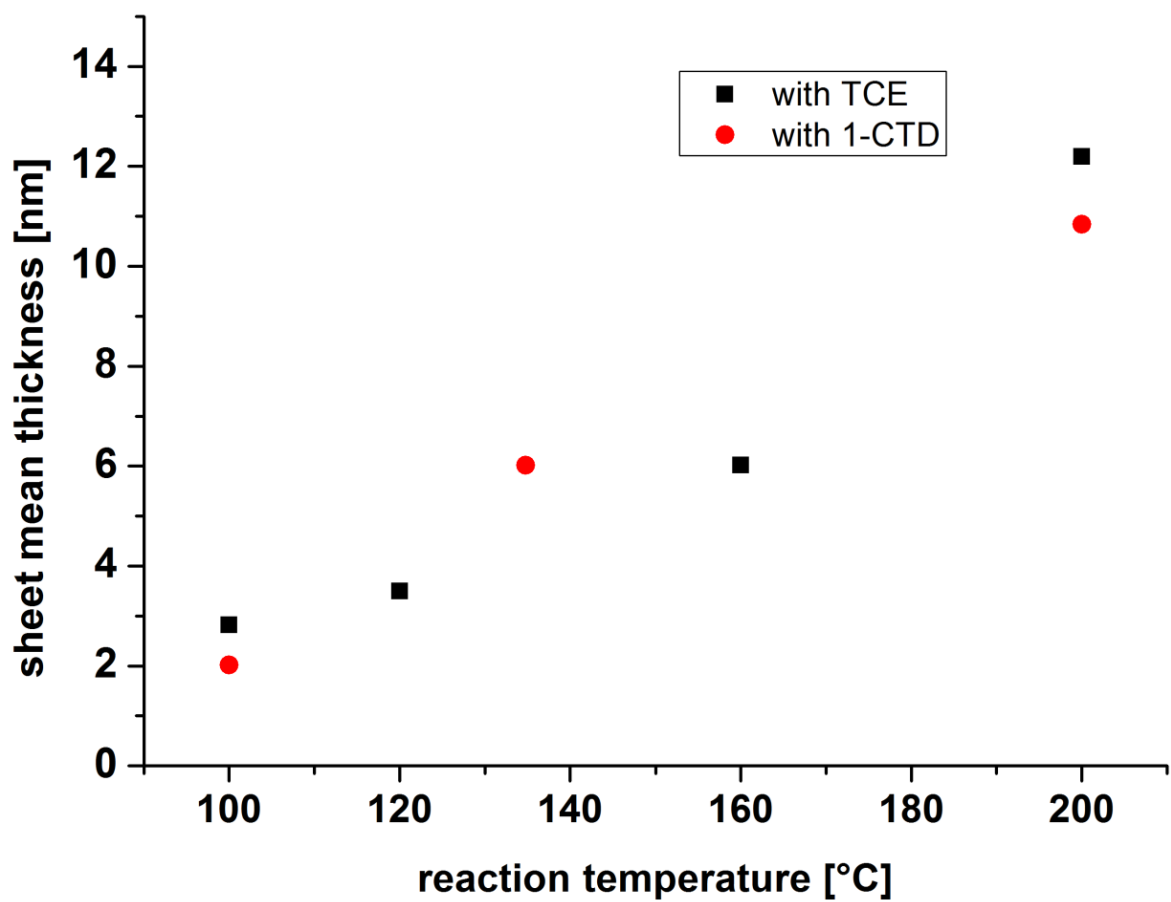

**Figure 6: Thickness of PbS sheets as a function of the reaction temperature.** *The black rectangles represent the results with TCE as co-ligand while the red circles represent the ones with 1-CTD. The thickness was calculated using the Scherrer's equation[49] by fitting the (200) XRD peak of the lead sulfide sheets. Despite their different structure both chloroalkanes produce nanosheets with comparable thickness at similar reaction temperatures (Fig. 5 and Fig. S4). With elevated temperature the sheets become thicker.*